\documentclass[aps,prl,twocolumn,nofootinbib,superscriptaddress,preprintnumbers,longbibliography]{revtex4-2}
\usepackage[utf8]{inputenc} 
\usepackage[T1]{fontenc}
\usepackage{ucs}
\usepackage{epsfig}
\usepackage{graphicx}
\usepackage[english]{babel}
\usepackage{hyphenat}
\usepackage{amsmath}
\usepackage{amssymb}
\usepackage{bbold}
\usepackage{mathtools}
\usepackage{mathrsfs}
\usepackage{slashed}
\usepackage{epstopdf}
\usepackage[dvipsnames]{xcolor}
\usepackage{braket}
\usepackage{booktabs}
\definecolor{lcolor}{rgb}{0.,0.0,0.}
\definecolor{citcolor}{rgb}{0,0.,0.5}
\usepackage[breaklinks,colorlinks,urlcolor=blue,citecolor=blue,linkcolor=blue]{hyperref}
\usepackage{multirow}
\usepackage{ltablex}
\usepackage{soul}

\usepackage[e]{esvect}

\usepackage{media9}

\def\cO{{\cal O}}

\newcommand{\secn}[1]{Section~1}
\newcommand{\appn}[1]{Appendix~1}

\long\def\comment#1{ }

\def\and{\quad\text{and}\quad}

\def\0{{\boldsymbol 0}}
\def\1{{\boldsymbol 1}}
\def\p{{\boldsymbol p}}

\def\r{{\boldsymbol r}}

\def\v{{\boldsymbol v}}

\def\b{{\boldsymbol b}}
\def\0{{\boldsymbol 0}}

\def\P{{\boldsymbol P}}

\def\bn{{\boldsymbol n}}

\newcommand{\jewel}{\textsc{Jewel}}

\renewcommand\o{\omega}

\renewcommand{\part}{{\rm part}}

\newcommand{\be}{\begin{equation}}
\newcommand{\ee}{\end{equation}}
\newcommand{\bes}{\begin{subequations}}
\newcommand{\ees}{\end{subequations}}
\newcommand{\bea}{\begin{eqnarray}}
\newcommand{\eea}{\end{eqnarray}}

\newcommand{\nn}{\nonumber \\}

\def\bea#1\eea{\begin{align}#1\end{align}}
\newcommand{\bef}{\begin{figure}[h!tb]\centering}
\newcommand{\eef}{\end{figure}}

\allowdisplaybreaks

\newcommand\n{{\bf n}}
\renewcommand\b{{\bf b}}

\begin{document}

\title{Early-Time Dynamics of Heavy-Ion Collisions through Energy Correlators: \\ {\small celestial blocks and the spacetime structure of out-of-equilibrium QCD matter}
}

\author{Jo\~{a}o Barata}
\email{joao.lourenco.henriques.barata@cern.ch}
\affiliation{CERN, Theoretical Physics Department, CH-1211, Geneva 23, Switzerland}

\author{José Guilherme Milhano}
\email[]{gmilhano@lip.pt}
\affiliation{LIP, Av. Prof. Gama Pinto, 2, P-1649-003 Lisboa, Portugal}
\affiliation{Departmento de Fisica, Instituto Superior Tecnico (IST), Universidade de Lisboa, Av. Rovisco Pais 1, P-1049-001 Lisboa, Portugal}

\author{Andrey V. Sadofyev}
\email{andrey.sadofyev@ehu.eus}
\affiliation{Department of Physics, University of the Basque Country UPV/EHU, P.O. Box 644, 48080 Bilbao, Spain}
\affiliation{IKERBASQUE, Basque Foundation for Science, Plaza Euskadi 5, 48009 Bilbao, Spain}
\affiliation{LIP, Av. Prof. Gama Pinto, 2, P-1649-003 Lisboa, Portugal}

\author{Jo\~{a}o M. Silva}
\email[ ]{joao.m.da.silva@tecnico.ulisboa.pt}
\affiliation{Departamento de Física Teórica y del Cosmos, Universidad de Granada, Campus de Fuentenueva,
E-18071 Granada, Spain}
\affiliation{LIP, Av. Prof. Gama Pinto, 2, P-1649-003 Lisboa, Portugal}
\affiliation{Departmento de Fisica, Instituto Superior Tecnico (IST), Universidade de Lisboa, Av. Rovisco Pais 1, P-1049-001 Lisboa, Portugal}

\preprint{CERN-TH-2025-257}

\begin{abstract}
Ultrarelativistic heavy-ion collisions provide a unique window into far-from-equilibrium states of QCD matter. 
The initial stages of these events are characterized by highly anisotropic, nonthermal dynamics that precede hydrodynamization, yet they remain largely inaccessible through conventional soft observables. In this work, we show that the substructure of mid-rapidity jets provides direct sensitivity to the spacetime structure of this early, anisotropic phase.
Using classical Yang–Mills simulations and effective kinetic theory to model the early-time evolution of the jet quenching parameter, we compute the azimuthally differential energy–energy correlator within the BDMPS-Z framework. By decomposing the result into celestial blocks, we isolate the coefficients that encode the anisotropic geometry and dynamics of the underlying medium. We identify an observable that couples directly to spatial anisotropies in the out-of-equilibrium QCD matter and also discuss the impact of medium response on its behavior.
We further extend our study to mid-rapidity jets generated with the \jewel\ Monte-Carlo, adjusted to incorporate an anisotropic medium background, and find qualitative agreement with the analytical expectations. Finally, we discuss how higher-point energy correlators and generalized energy-flow operators can enhance the sensitivity to the microscopic structure of far-from-equilibrium QCD matter.
\end{abstract}

\maketitle

\section{Introduction}\label{sec:intro}
The early stages of heavy-ion collisions (HICs) are characterized by an evolution from highly anisotropic, far-from-equilibrium initial conditions toward a nearly equilibrated state that precedes the hydrodynamic expansion of the quark–gluon plasma (QGP)~\cite{Berges:2020fwq,Schlichting:2019abc}. These initial stages offer a unique opportunity to investigate the many-body out-of-equilibrium dynamics of QCD, a regime that is difficult to reproduce in other experimental settings. 
Thus, HIC experiments might provide invaluable insight into the behavior of gauge theories under extreme conditions and their relaxation toward equilibrium, which remains one of the least understood frontiers of quantum field theory.

\begin{figure}[ht!]
    \centering
    \includegraphics[width=0.8\columnwidth]{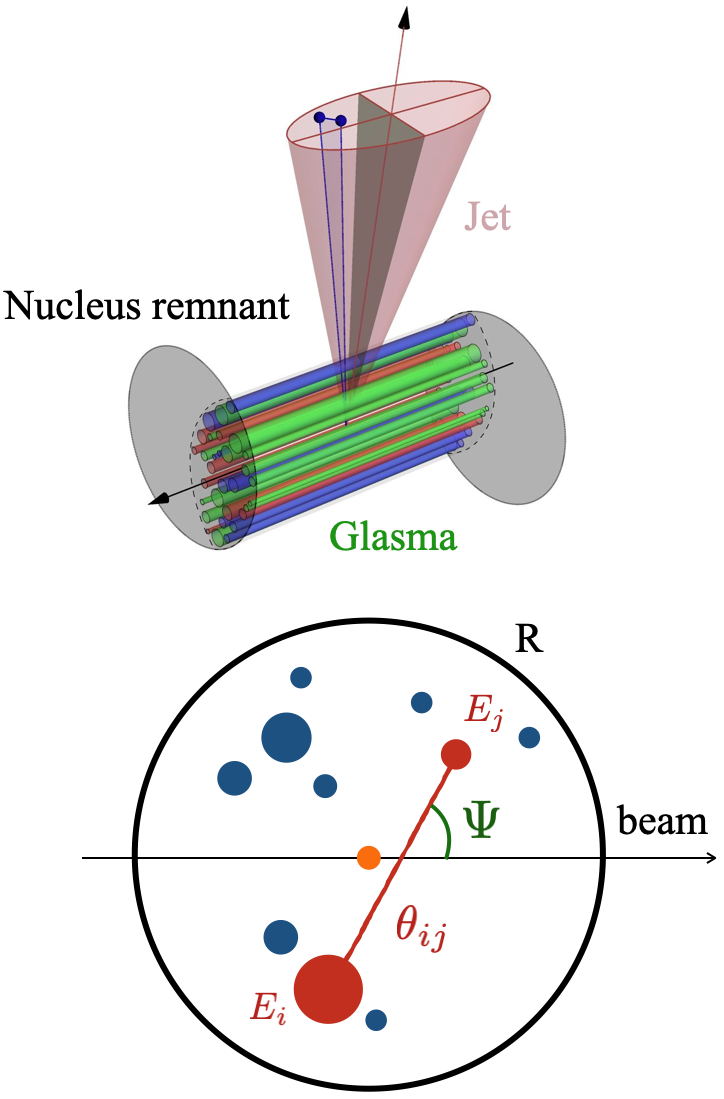}
    \caption{\textbf{Top:} Illustration of a typical mid-rapidity jet configuration considered in this work. Elongated color tubes represent the initial Glasma configuration which dominate the first instances after the collision, and are described within classical Yang-Mills theory. In blue we show one possible particle pair contributing to the EEC observable. The elongated jet profile along the beam indicates the expected deformation in the energy distribution inside the jet. \textbf{Bottom:} Particle based in-jet EEC. Here we indicate the relative angle $\theta_{ij}$, between a particle pair with respective energies $E_i$, $E_j$. The azimuthal angle $\Psi$ is defined, for each pair, with respect to the beamline. The jet radius is $R$ and the orange circle denotes its axis.}
    \label{fig:Glasma}
\end{figure}

Despite the clear theoretical interest in exploring the early stages of HICs, so far very limited experimental data are available that constrain or inform about the properties of such states. This is natural, since most measurements focus on the hadronic remnants of the bulk matter; as this hydrodynamically evolves for a long time, memory of the initial stages is nearly lost (see e.g.~\cite{Liu:2015nwa} for further discussion). Any hope of probing the out-of-equilibrium dynamics in HICs must lie in observables sensitive to local properties of the early stages that do not fully thermalize.

In this respect, there has been a recent dedicated effort toward using hard probes, such as jets and heavy flavor, to exactly fulfill this role. In the case of jets, this has been motivated by several theoretical developments. In particular, recent considerations of the jet quenching transport coefficient $\hat q$ in the Glasma and kinetic theory stages~\cite{Ipp:2020mjc,Ipp:2020nfu,Avramescu:2023qvv,Carrington:2020sww,Carrington:2021dvw,Hauksson:2021okc,Boguslavski:2024ezg,Boguslavski:2024jwr,Avramescu:2025lhr,Altenburger:2025iqa} have shown that its absolute value can be significantly larger than what is expected for the QGP phase. Moreover, it has been found that $\hat q$ becomes anisotropic, leading to a non-trivial azimuthal structure for the radiation inside the jet cone. These developments on the description of transport in the bulk have been accompanied by an extension of the jet quenching toolkit to 
the case of structured and evolving matter~\cite{Sadofyev:2021ohn,Sadofyev:2022hhw,Andres:2022ndd,Barata:2022krd,Fu:2022idl,Barata:2022utc,Hauksson:2023tze,Barata:2023qds,Kuzmin:2023hko,Barata:2024bqp,Kuzmin:2024smy,Barata:2024xwy,Barata:2025htx,Sengupta:2025kdr,He:2020iow,Silva:2025dan,Barata:2025agq}, and the development of observables which can directly probe such structures, both in the transverse direction to the jet axis and along the longitudinal (time) direction~\cite{He:2020iow, Apolinario:2020uvt, Antiporda:2021hpk, Barata:2023zqg, Bahder:2024jpa, Apolinario:2024hsm, Ke:2024emw, Barata:2025uxp,Apolinario:2020uvt}, see also~\cite{Adhya:2024nwx,Barata:2024xwy, Pablos:2025cli,Barata:2025agq} for recent dedicated discussions on the effects of the early stages in jet observables. Although several questions remain open, the combination of all these efforts marks a clear path toward a more differential treatment of jet substructure in the presence of increasingly more complex matter backgrounds. It is worth mentioning that this effort also provides theoretical backing towards a jet substructure program in smaller collisional systems~\cite{Grosse-Oetringhaus:2024bwr}, where the relevance of out-of-equilibrium matter states can be more evident.

In this work, we discuss how the imprint of the initial stages of HICs can be recovered by measuring the azimuthal structure of radiation inside mid-rapidity jets. Expanding on~\cite{Barata:2023zqg}, we argue that multi-point correlators of energy flow operators 
\begin{align}\label{eq:def_Ec}
 \mathcal{E}(\bn) = \lim_{r\to \infty} r^{2
 } \int_0^\infty dt\, n^i T^{0i}(\bn)\, ,
\end{align}
are natural objects to probe the detailed structure inside jets and can be
connected to the spatial structure of the initial-state matter. This reasoning is illustrated in Fig.~\ref{fig:Glasma} (top): a jet produced at mid-rapidity will preferentially exhibit an elongated energy distribution along the beam-axis due to the anisotropic character of the underlying matter. The latter is shown here for the earliest times, where a classical Yang-Mills Glasma initial state is expected to emerge~\cite{Iancu:2003xm,Lappi:2006fp,Gelis:2010nm,Gelis:2012ri,Gelis:2015gza}, but the construction also applies to the subsequent stages often described in the context of effective kinetic theory (EKT) of QCD~\cite{Arnold:2002zm}. Thus, measuring correlation functions of $\mathcal{E}(\n)$ inside jets, differentially in both the longitudinal and azimuthal angles, would provide a way to characterize the spatial structure of the bulk. In this work, we focus on the two-point function, the energy-energy correlator (EEC), whose double-differential form has been recently considered in other related contexts~\cite{Song:2025bdj}. Although the spacetime structure can be reconstructed from higher point correlation functions, their computation and measurement in HICs becomes rather involved~\cite{Budhraja:2025ulx,Barata:2025fzd,Ananya:2025qaf,Bossi:2024qho}.

To further clarify the logic of our basic arguments, let us comment on how the background nuclear matter evolution affects in-jet correlators of the energy flow operator $\mathcal{E}(\bn)$, e.g. $\langle X|  \mathcal{E}(\bn_1) \mathcal{E}(\bn_2)|X\rangle$ with $|X\rangle$ representing some out-of-equilibrium state in the \textit{microscopic} theory which also includes a jet-like excitation on top of the bulk. One may regard the medium as an underlying time-dependent filter acting on the jet: while the medium contribution to the EEC depends only on the final state, the in-jet EEC is sensitive to the entire history of jet-matter interactions. However, certain features of those interactions only depend on particular stages of the evolution. This observation can be illustrated with the following toy model: consider a single spin, serving as a probe, immersed in a time-dependent background magnetic field. The averaged spin plays the role of a single-point energy correlator, gauging the anisotropy of the probe-matter interactions, and can be written as
\begin{align}
    \mathcal{S}_z=\frac{1}{T} \int_0^T dt\, \left\langle s_0\left|  \hat{S}_z(t)\right| s_0\right\rangle\, ,
\end{align}
where $|s_0\rangle$ specifies the initial state of the system. If we now assume that the field is nearly constant for some initial time interval $t\in(0,t_0)$ and rapidly fluctuating for $t\in(t_0,T)$, then the contribution of the latter regime can be neglected. Thus, the observable is determined by the properties of the bulk matter prior to the $t_0$ scale.

Translating this simpler analogy to our context, as long as strong directional effects along the beamline dominantly characterize the initial stages, the contributions associated with the hydrodynamic QGP phase are expected to be subdominant. Moreover, at mid-rapidity, flow velocity and temperature gradients along the beam direction can be neglected for central collisions.
Consequently, the flow and gradients can be modeled as mainly radial for our proof-of-concept considerations. Thus, assuming the jet to be produced sufficiently close to the center of the matter, we neglect hydrodynamic effects in jet quenching, leaving them for future study.

This picture is, in essence, analogous to that of the \textit{fireball} state considered in~\cite{Delacretaz:2018cfk}. Extending our model picture to HICs, we can write
\begin{align}
\langle\mathcal{E}\rangle\simeq \sum_{\text{initial}}\int_0^{t_{0}} dt\, \langle \text{matter}(t)| \mathcal{E}_{\text{Schr}}| \text{matter}(t)\rangle\,,
\end{align}
where $t_0$ should be understood as a characteristic isotropization time, closely related to the hydrodynamization time~\cite{Kurkela:2015qoa,Baier:2000sb}. This illustration highlights how the anisotropy of the initial stage is imprinted in the jet substructure. While the structure of asymptotic energy flows in HICs is far more complex than this simple example~\cite{Andres:2024ksi,Yang:2023dwc,Barata:2023bhh,Singh:2024vwb,Xing:2024yrb,Apolinario:2025vtx, Liu:2025ufp,Barata:2024wsu,Ke:2025ibt}, we will show that, with a non-trivial anisotropic structure for times $t<t_0$, a similar form emerges for a constrained collection of jets.
We also note that the present analysis is, in our view, more challenging to carry out if considering jet shapes as, in that case, a \textit{practical} notion of time~\cite{Apolinario:2024hsm, Apolinario:2020uvt} is needed, making the observable structure more complex.

This paper is organized as follows. We first discuss the structure of the EEC in the OPE limit in the presence of a vector excitation. We then introduce a simple EEC-based observable that directly targets contributions sensitive to the azimuthal structure of the state. 
Following this, we show that, within a simple analytical model, the same structure is expected for the EEC of jets evolving in the early stages of HICs, and compute the observable using benchmark values of the jet quenching parameter $\hat q$ taken from state-of-the-art numerical simulations, for illustrative purposes.
We then extend the discussion to a Monte Carlo–based calculation with a simple model for an anisotropic background, as a further illustration, where a similar imprint of the bulk geometry is found in final-state observables.
We conclude by summarizing the main results and elaborating on possible connections between 
azimuthally differential EECs and higher-point energy correlators~\cite{Chen:2019bpb}, generalized detectors~\cite{Korchemsky:2021okt,Caron-Huot:2022lff}, and Glasma-state correlators~\cite{Krasnitz:1998ns}. Some of the details of the considerations presented here are omitted for brevity and given in the Appendix.

\section{Celestial blocks and the OPE limit}\label{sec:Celestial blocks and the OPE limit}
Energy flow operators provide a powerful tool to characterize the substructure of jets, see~\cite{Moult:2025nhu} for a recent review. In particular, in the conformal limit of QCD, the form of the associated correlation functions can be partially determined from symmetry arguments, allowing to isolate cleanly the dynamical elements. Although in the HICs context such arguments are insufficient for a full characterization of the events, energy flows are still an attractive tool due to their theoretical simplicity that might allow for a more direct connection to experiment.

Here we want to characterize the EEC measured inside jets produced in the presence of an initially highly anisotropic state along the beam-axis as in Fig.~\ref{fig:Glasma}. We recall first the structure of the celestial block decomposition of the EEC in the presence of a vector perturbation acting in a trivial state, which was first studied in detail in~\cite{Kologlu:2019mfz} and more recently applied to hadronic collisions~\cite{Chen:2025rjc}; see also~\cite{Hofman:2008ar, Chang:2022ryc,Chen:2022jhb,Chang:2020qpj} for related discussions. The EEC double differential distribution is defined as
\begin{widetext}
    \begin{align}\label{eq:main1}
\frac{1}{\sin\chi\sin\Psi}
\frac{d\Sigma}{d\chi d\Psi
}\Bigg|_{\eta=0}&=\int d\n_1 d\n_2 \, \frac{\langle\mathcal{E}(\n_1)\mathcal{E}(\n_2)\rangle}{p_t^2}
\, \delta(\n_1\cdot\n_2-\cos\chi)\, \delta\left(\frac{\n_1-\n_2}{|\n_1-\n_2|}\cdot\b-\cos\Psi
\right)
\,,
\end{align}
\end{widetext}
where $p_t$ is the overall large energy scale we identify with the measured jet energy.\footnote{For the purposes of this work, we neglect energy loss effects, which would not qualitatively alter our conclusions.} Anticipating the physical application of interest, here we have projected the azimuthal angle $\Psi$ such that the reference direction $\b$ is set by the beam, see Fig.~\ref{fig:Glasma} (bottom). The angle $\chi$ measures the angular distance between the energy flow operators. We note that compared to the previous study in~\cite{Barata:2023zqg}, where the reference direction for the azimuthal angle is never specified, here the particular structure of the initial stages, preferentially aligned along the beam direction, results in a \textit{natural} reference for the azimuthal correlations. In the collinear limit, $\chi \ll1$, the angle $\Psi$ can be interpreted as the azimuthal angle around the jet axis, while $\eta$ is the jet's pseudorapidity.
We will focus on jets at mid-rapidity, $\eta=0$, corresponding to rapidity $Y=0$; being understood that this 
corresponds to jets reconstructed in the central rapidity bin. It is also instructive to consider a particle description for Eq.~\eqref{eq:main1}, suitable for experimental applications
\begin{align}\label{eq:main2}
\frac{1}{\sin\chi\sin\Psi}\frac{d\Sigma}{d\chi d\Psi}\Bigg|_{\eta=0}
&=\sum_{i\neq j}\frac{E_iE_j}{p_t^2} \delta(\cos\theta_{ij}-\cos\chi) \nn 
&\hspace{-0 cm}\times \delta\left(\frac{\theta_{jb}-\theta_{ib}}{\theta_{ij}}-\cos\Psi
\right)\,,
\end{align}
where $\theta_{ij}\ll1$ is the relative angle between the particle pair $\{i,j\}$,
while $\theta_{ib}$ is the angle between particle $i$ and the beam direction.
In the collinear limit, and for mid-rapidity jets, 
one may readily notice that $|\pi/2-\theta_{ib}|\ll1$. 

For a vector perturbation, the two-point correlator admits the celestial block 
decomposition~\cite{Kologlu:2019mfz,Chen:2025rjc}:
\begin{align}
\langle \mathcal{E}(\bn_1) \mathcal{E}(\bn_2)\rangle = \frac{2 p_t^2}{(1-\n_1\cdot\n_2)^3}\sum_{\delta,j}\int_\gamma \frac{ c_{\delta,j,\gamma}}{2\pi i } w^\gamma G_{\delta ,j,\gamma}(z,\bar z)  \, .
\end{align}
The celestial blocks are given in terms of hypergeometric functions: 
\begin{align}
  G_{\delta ,j,\gamma}(z,\bar z)  &= z^h \bar z^{\bar h} \mathcal{F}(h,\gamma,z) + \bar z^{ h} z^{\bar h} \, \mathcal{F}(h,\gamma, \bar z)  \, ,
\end{align}
with $\mathcal{F}(h,\gamma,z)= \, _2F_1(h,h-\gamma,2h,z) \,  _2F_1(\bar h, \bar h-\gamma,2 \bar h, \bar z) $.
Here, $\delta=\tau+2$ is the celestial dimension directly connected to the twist $\tau$, the transverse spin satisfies $j\leq \tau-2$, and $2h=\delta -j$, $2 \bar h= \delta + j$. The kinematical variables $z$, $\bar z$
are defined in terms of the \textit{collider} coordinates as $z\bar z = 1-2 e^{\Delta Y} \cos(\Delta \Phi) + e^{2\Delta Y}$, and $(1-z)(1-\bar z) = e^{2\Delta Y}$,
where $\Delta Y$ and $\Delta \Phi$ denote, respectively, the rapidity and azimuthal separation of the detectors, defined with respect to the beam direction. We are interested in the collinear limit where $|z|\to 0$ and consider the detectors to be close to mid-rapidity. Parametrizing $z = r e^{i\Psi}$ and expanding the hypergeometric functions, we readily find $G_{\delta ,j,\gamma}(z,\bar z) \simeq  2 r^{\tau+2} \cos\left( \Psi j\right)$, at leading power in $r$. One should note that this expansion is kinematic in nature for the fixed global spin $J=3$ basis and therefore persists in generic interacting theories.
Noticing that at mid-rapidity one can take the limit $\gamma \to 0$, we can directly perform a Mellin transform extracting the residue from the pole; inserting the necessary Jacobians, we can write the double differential EEC in the OPE limit as 
\begin{align}\label{eq:EEC_OPE}
  \frac{ d\Sigma}{d\chi d\Psi}  = \sum_{k=1} \left[\sum_{j=0,2,4
  }^{\tau-2} c_{\tau,j}\,  \chi^{\tau-3} \cos\left( \Psi j\right)\right]_{\tau=2k} \, .
\end{align}
Notice that in a conformal setting, the global and transverse spins are constrained by the relation $J \ge 1 + j/2$. 
It is thus instructive to expand Eq.~\eqref{eq:EEC_OPE} and combine it with the contributions associated with medium-induced effects:
\begin{widetext}
\begin{align}\label{eq:general_EEC}
              {\hspace{-.8 cm}}\frac{ d\Sigma}{d\chi d\Psi}  &= \underbrace{{\color{blue}\sum_{k=1} \left(c_{\tau,0}+b_{\tau>2,0}+a_{\tau> 2,0}\right)  \chi^{\tau-3}} + \sum_{k=2} \left(c_{\tau,2}+b_{\tau>4,2}+{\color{red} a_{\tau>4,2}}\right)\chi^{\tau-3} \cos(2\Psi)+ \sum_{k=3} (c_{\tau,4}+b_{\tau>6,4}+{\color{red}a_{\tau>6,4}}) \chi^{\tau-3} \cos(4\Psi)}_{\mathcal{O}_{J=3}} \nn 
              &+ \,  [\mathcal{O}^{j=4}_{J\geq 5}] \, , \quad {\rm with}\,\,\,\, \tau=2k\, .
    \end{align}    
\end{widetext}
In Eq.~\eqref{eq:general_EEC} the terms in blue were discussed in~\cite{Barata:2025fzd} in the HICs' context; they correspond to the \textit{vacuum} contribution $c_\tau$, the \textit{higher-twist} terms $b_\tau$ enhanced by the nuclear environment, and the terms associated with the (isotropic) classical medium response $a_\tau$. The novel contributions $c/b_{\tau,(2,4,\cdots)}$ associated to the leading global spin ($J=3$) operators match the ones first identified in the celestial OPE for a vector perturbations in~\cite{Kologlu:2019mfz}.  
The remaining terms (in red) correspond to the azimuthally asymmetric classical contributions connected to uncorrelated (classical) flows emerging from the response of the bulk matter. In the second line, we denote the contributions associated with higher-spin light-ray operators ($J\geq 5$) acted upon by appropriate differential operators, which generate higher transverse-spin structures within the fixed $J=3$ sector, in agreement with the established light-ray OPE formula for $J=3$ and its longitudinal descendants with $j=4$~
\cite{Chang:2020qpj}. We note that medium induced terms could lead to $j>4$ contributions, which  are expected to be harder to access experimentally.

While the vacuum contributions at different transverse spins $j$ are constrained by the unitarity bound $\tau \ge j+2$, the stricter constraints satisfied by the $a_{\tau,j}$ and $b_{\tau,j}$ terms are not implied by the celestial block (light-ray OPE) expansion alone. Rather, they emerge from explicit dynamical considerations, as we show below\footnote{This specific pattern of medium-induced terms may in fact be more general and potentially model-independent, possibly arising already at the level of the light-ray OPE. We leave a detailed investigation of this possibility for future work.} and as also discussed in~\cite{Barata:2025fzd}.

Having presented the structure of the double-differential EEC distribution in the collinear limit, we now discuss how it can be used to isolate directional effects which can couple to the matter background. To that end we introduce the \textit{clover} EEC, $d\Sigma^{\Delta}$, defined in terms of the azimuthally integrated distribution
\begin{align}\label{eq:EEC_clover_def}
  \frac{d\Sigma^\Omega}{d\chi} = \int_\Omega d\Psi \frac{d\Sigma}{d\chi d\Psi}  \, ,
\end{align}
where $\Omega$ defines a region inside the jet azimuthal plane. In what follows, we consider a horizontal ($h$) clover $\Psi \in (-\pi/4; \pi/4)$ and a vertical ($v$) clover $\Psi \in (\pi/4,3 \pi/4)$, where the angle is defined with respect to the beam line.

We define the clover distribution as the difference of the previous two: $\Sigma^{\Delta}\equiv\Sigma^{v}-\Sigma^{h}$, such that it does not depend on leading spin terms. In what follows we discuss how this structure maps to jet quenching observables, and show that in a boost-invariant initial state, the form of the leading-order differential EEC directly connects to Eq.~\eqref{eq:general_EEC}.

\section{Jet evolution in a boost invariant initial state}\label{sec:Jet evolution in a boost invariant initial state}

We now study how the \textit{clover} EEC behaves for jets propagating through matter produced in HICs. This matter starts far-from-equilibrium and, assuming a weak-coupling picture, its earliest stages are commonly first described by a nearly classical highly occupied gluonic state, the Glasma \cite{McLerran:1993ni,Kovner:1995ts,Krasnitz:1999wc,Lappi:2006fp}, followed by a pre-equilibrium evolution typically discussed within the context of kinetic theory~\cite{Baier:2000sb,Berges:2013eia,Kurkela:2015qoa}. In the former regime, the chromoelectric and chromomagnetic fields are predominantly longitudinal, aligned with the beam axis, and this is a direct manifestation of the initial anisotropy of the system. This anisotropy persists throughout the pre-equilibrium evolution until the system approaches local equilibrium. At this point, it becomes nearly isotropic and ultimately it hydrodynamizes and thermalizes, see e.g.~\cite{Schlichting:2019abc} for a recent review. Consequently, one expects jet substructure to develop differently along the beam direction and transverse to it.

We now proceed to show how initial-stage (IS) anisotropies imprint themselves on jet substructure, through both the medium-induced radiation spectrum and the medium response to the jet, and how these effects can be connected to the EEC decomposition. We recall first the result discussed in~\cite{Barata:2023zqg,Barata:2025uxp,Barata:2024bqp} for the production of radiation in a transversely anisotropic plasma in terms of the respective EEC distribution: 
\begin{align}
    \frac{d\Sigma_{\rm IS}}{d\chi d\Psi} = \frac{1}{2\pi}\frac{d\Sigma_{\rm IS}}{d\chi } + \sum_{n=1}^{\infty} v^{\rm EEC}_{2n}(\chi)  \cos(2n \Psi) \, .
\end{align}
As noted in~\cite{Barata:2024bqp}, the presence of the even cosine Fourier series for mid-rapidity jets is constrained solely by the symmetries under $\Psi \rightarrow \Psi \pm \pi$ and $\Psi \rightarrow -\Psi$, much like in the celestial block structure above.\footnote{One should note that these symmetries can be broken, for instance by sensitivity to final-state spins and their correlations, see e.g.~\cite{Silva:2025dan}, or by stronger anisotropies of the medium, as occur in off-central collisions, resulting in a more generic Fourier series.} Averaging over spins and focusing on central collisions, one can write the \textit{clover} EEC as
\begin{align}
   \frac{d\Sigma^\Delta_{\rm IS}}{d\chi} =  \sum_{n=1}^{\infty} \frac{2}{n} \,  v^{\rm EEC}_{2n}(\chi) \, \sin\left(\frac{n\pi}{2}\right) \, ,
\end{align}
which implies only odd values of $n$ contribute. The Fourier coefficients $v_{2n}^{\rm EEC}$ admit a \textit{twist-like} expansion which allows them to be directly mapped to the OPE coefficients derived above, identifying $j=2n$. 
One may further note that the specific projection in Eq.~\eqref{eq:main1} (see Appendix) enforces the expansion
\begin{align}\label{eq:v2n_twist_expansion}
    & v_{2n}^{\rm EEC}(\chi) = \sum_{k=n+2}v_{2n,2k}^{\rm EEC}\chi^{2k-3+s}\, , \quad n\geq 1\, ,
\end{align} 
thereby constraining the contributing twists to satisfy $\tau\geq j+4-s$, as long as the expansion in $\cos\Psi$ exists, e.g. for small anisotropy. Here, $s$ is a shift power assumed to be common for all harmonics and set by the potential collinear singularity (controlled by the short-distance behavior);  both medium-induced
jet modifications, in our treatment, and classical flows due to the medium response lead to $s=0$.
Interestingly, the expansion of the $2n$-th harmonic coefficient in powers of anisotropy strength $\xi$, e.g., in powers of the difference between diffusion constants along and orthogonal to the beamline, starts at power $\xi^n$.
Using this result and Eq.~\eqref{eq:v2n_twist_expansion}, the \textit{clover} EEC
reads
\begin{align}
    \frac{d\Sigma^\Delta_{\rm IS}}{d\chi} 
    = 2v_2^{\rm EEC}(\chi) + \cO(\xi^3\chi^7)\, ,
\end{align}
where the leading term in $v_2^{\rm EEC}$ is of order $\cO(\xi \, \chi^3)$.  
To determine the observable introduced above, one must fully characterize both the properties of the bulk medium and the jet--medium interaction, which are encoded in a medium-response model and in the medium-induced spectrum. In the absence of a phenomenologically reliable model for the medium response during the initial stages of the evolution, we will omit this contribution in our further estimates and focus instead on medium-induced modifications to the $g\rightarrow c\bar c$ cross-section.

The latter is provided in the Appendix for a generic expanding medium and determined by the jet quenching parameter $\hat q$ within the BDMPS-Z approximation~\cite{Baier:1996kr,Zakharov:1996fv}. In our model, the system is further characterized by an anisotropy along the beam direction, parametrized by the ratio of jet quenching parameters parallel and transverse to the beam, $\hat q_y/\hat q_z$. To capture the evolution of the medium, we use state of the art numerical results for the $\hat{q}$ extracted from classical statistical simulations of the earliest Glasma stage~\cite{Avramescu:2023qvv, Ipp:2020nfu} and the subsequent kinetic theory epoch~\cite{Boguslavski:2024ezg,Boguslavski:2024jwr}.\footnote{We note that these studies only explored the jet quenching parameter in the context of momentum diffusion. When discussing radiative processes, this $\hat q$ might not be the only relevant medium parameter, see~\cite{Barata:2024xwy} for an explicit example. However, in the analytic model used, based on the BDMPS-Z picture for energy loss, the radiative rate is again fully determined by the jet quenching parameter obtained from momentum broadening, see also the recent discussion in~\cite{Barata:2025agq}.} 
The $\hat q_i$ extracted from Glasma simulations are calculated as the $\tau$ derivative of the momentum broadening in the $i$th direction, which in turn is obtained from gauge invariant correlators of electric and magnetic fields evaluated along the jet trajectory. In these simulations the saturation momentum is fixed to $Q_s = 2$ GeV, roughly corresponding to the most central collisions at LHC energies. The EKT data is matched to the Glasma data by imposing the same energy density, such that one fixes $Q_s = 1.4$ GeV, see e.g.~\cite{Boguslavski:2024ezg}.

\begin{figure}[h!]
    \centering
    \includegraphics[width=1\columnwidth]{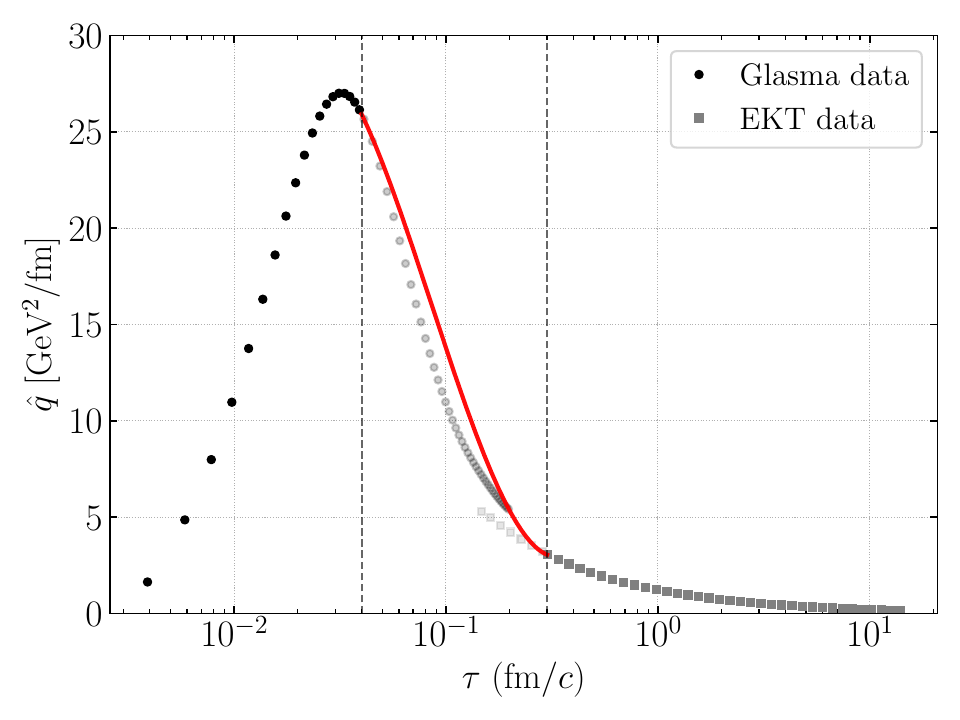}
    \includegraphics[width=1\columnwidth]{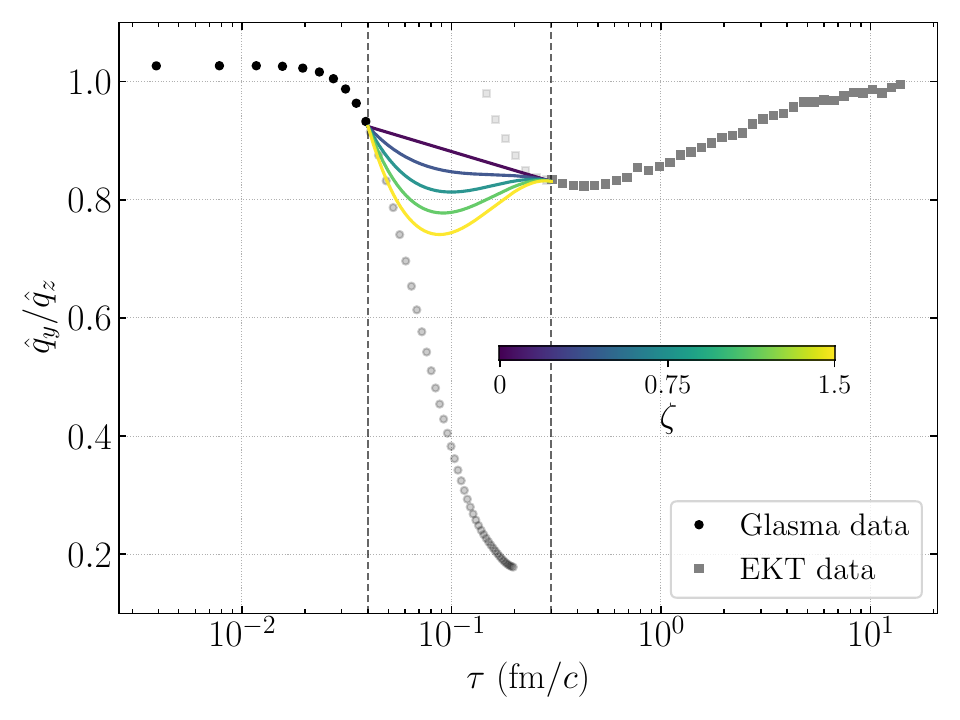}
    \caption{Scalar part of the (quark) jet quenching parameter $\hat q = \hat q_y + \hat q_z$ (\textbf{top} panel) and anisotropy ratio $\hat q_y/\hat q_z$ (\textbf{bottom} panel) as a function of light-cone time $\tau$. The series of curves in the transition region $\tau \in (0.04, 0.3)$ fm/c corresponds to a family of Hermite polynomials depending on the parameter $\zeta \in (0,1.5)$. For $\hat q$ we take the smoothest of these ($\zeta=1$).}
    \label{fig:qhat}
\end{figure}

In Fig.~\ref{fig:qhat} we show the evolution of $\hat q$ as a function of proper time.  
On the top panel we describe the trace of the transport coefficient, $\hat q = \hat{q}_z + \hat{q}_y$, while on the bottom panel we show the ratio of the transport coefficient evaluated along the beam-axis ($\hat{q}_z$) and in the transverse direction ($\hat{q}_y$). 
In both panels, we display the entire data set, which includes the region where both descriptions overlap as is parametrically excepted after $\tau\sim1/Q_s$.
The precise transition between these regimes is complex to (numerically) achieve in practice, even though such a connection is conceptually understood~\cite{Epelbaum:2013waa,Berges:2014yta,Romatschke:2005ag,Romatschke:2005pm,Romatschke:2006nk,Mrowczynski:1993qm}; currently, both sides are matched at $\tau \sim 1/Q_s$ by requiring the same averaged energy density.

Here we connect these two descriptions by building a cubic Hermite interpolant in the transition region.\footnote{Note that our approach here is rather practical, and we do not discuss the dependence on, e.g., the choice of saturation scale, which can change the values of $\hat q$. Although this would alter the values of the extracted observables, such modifications would not affect the conclusions of our study at a qualitative level.} To do so, we truncate the data set for the evolution in Glasma at $\tau_1 =0.04$ fm/c and include the data from EKT after $\tau_2 \equiv \tau_{\rm EKT} =0.3$ fm/c. We then extract the 
behavior of each regime near these stitching points by performing local least--squares 
polynomial fits (up to cubic order) to the Glasma data for $\tau < \tau_{1}$ and to the 
EKT data for $\tau > \tau_{2}$. These fits provide both the interpolated values 
$y_{1}, y_{2}$ and their local slopes $y_{1}', y_{2}'$ at the boundaries. The transition 
region is then parameterized in terms of the log--time variable 
$s = (\ln \tau - \ln \tau_{1})/(\ln \tau_{2} - \ln \tau_{1}) \in [0,1]$, and the bridge is defined using a cubic Hermite interpolant whose 
endpoint slopes are blended with the secant slope through a tunable parameter 
$\zeta$, such that $\zeta = 0$ yields a straight line connection while 
$\zeta = 1$ recovers the natural slope from the interpolated data. The Hermite interpolant polynomial then reads
\begin{align}
\label{eq:hermite-bridge}
p_\zeta(\tau)&=
y_{1} h_{00}(s)
+ y_{2} h_{01}(s)\nn
&{\hspace{-.5 cm}}+ (\ln \tau_{2} - \ln \tau_{1})
\left[
m_1(\zeta)h_{10}(s)
+ m_2(\zeta) h_{11}(s)
\right],
\end{align}
where $h_{ij}(s)$ are the cubic Hermite basis functions in the interval $s\in(0,1)$ and $m_i(\zeta)=(1-\zeta)\,m_{\rm sec}+\zeta\,(t_{i} y_{i}')$, with secant slope $m_{\rm sec}=(y_2-y_1)/(\ln \tau_{2} - \ln \tau_{1})$. The family of curves $p_\zeta(\tau)$ is shown for a few values of $\zeta\in(0,1.5)$ for the ratio $\hat q_y / \hat q_z$ (lower panel) and for the scalar part $\hat q = \hat q_y + \hat q_z$ (upper panel) we took $\zeta = 1$, i.e., the curve with continuous derivative at the endpoints.

\begin{figure}[h!]
    \centering
    \includegraphics[width=1\columnwidth]{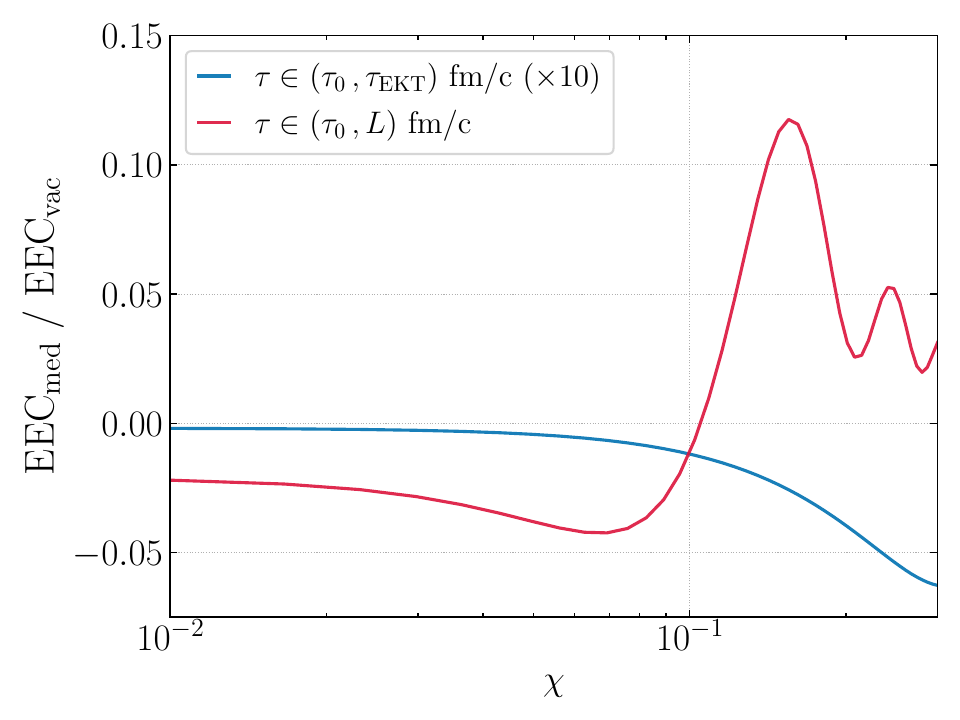}
    \includegraphics[width=1\columnwidth]{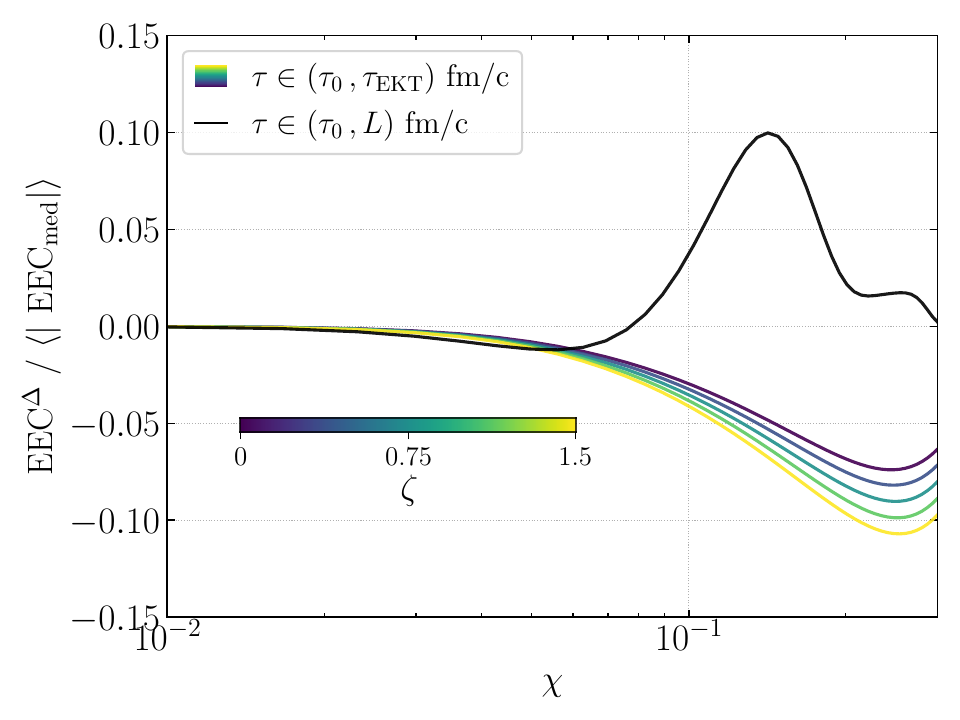}
    \caption{\textbf{Top}: Ratio between the $\Psi$ integrated medium contribution to the energy-energy correlator (EEC$_ {\rm med}$) and the vacuum EEC$_{\rm vac}$, for the $g\rightarrow c\bar c$ splitting . \textbf{Bottom}: Ratio between the \textit{clover} EEC and the $\chi-$averaged absolute value of EEC$_{\rm med}$ in the range $\chi\in(0.01,0.3)$, for the $g\rightarrow c\bar c$ splitting . The denominator is divided by $4$ to appropriately normalize it to half an hemisphere. We plot these quantities for initial time $\tau_0 \simeq 0.004$ fm/c and for two disparately different medium sizes --  $\tau_{\rm EKT} = 0.3$ fm/c and $L = 3$ fm/c.} 
    \label{fig:clover_eec}
\end{figure}

In Fig.~\ref{fig:clover_eec}, we show, in the top panel, the modification to the $g\rightarrow c\bar c$ EEC due to the medium-induced spectrum, while the bottom panel informs about the resulting azimuthal deformation. The two time intervals correspond to disparately different medium sizes -- one includes only the Glasma data and the interpolated transition region ($\tau \in (\tau_0, \tau_{\rm EKT})$, blue curve multiplied by $10$ for visualization), with $\tau_0 \simeq 0.004$ fm/c and $\tau_{\rm EKT} = 0.3$ fm/c and the other includes all regions ($\tau \in (\tau_0,L)$, red curve), with EKT data truncated at $L = 3$ fm/c. More specifically, the top panel shows the leading order result for the ratio between the pure medium contribution to the $\Psi$ integrated EEC (EEC$_{\rm med}$) and the $\Psi$ integrated vacuum EEC$_{\rm vac}$. Note we only plot a single curve, despite there being a family of curves for the anisotropy ratio in the bottom panel of Fig~\ref{fig:qhat}, since the $\Psi$-integrated EEC shows little sensitivity to the transition region. At the earliest stage (blue curve) the medium modification is much smaller (on average by a factor $\sim 20$ for large $\chi$) than the red curve, which includes the EKT data. Further, it is only non-negligible for sufficiently large $\chi$ and always negative. As for the red curve, which includes the EKT data, there is an average magnitude for medium modifications of a few percent. The non-vanishing small-$\chi$ behavior is expected due to an upper bound on formation time set by the mass of the heavy $c\bar c$ pair in the final state, as discussed in~\cite{Attems:2022otp,Attems:2022ubu,Barata:2025uxp}. For larger $\chi$ one sees an enhancement of the EEC, which is characteristic of medium modifications to energy-energy correlators. Note that the oscillations are not numerical in origin, but rather a consequence of the exponential term in Eq.~\eqref{eq:spectrum_aniso}.

In the bottom panel of Fig.~\ref{fig:clover_eec}, we show the ratio between the clover EEC$^\Delta$ for $g\rightarrow c\bar c$ 
and the average value of $|$EEC$_{\rm med}|$ in the range $\chi\in(0.01,0.3)$, such that one can gauge the magnitude of the directional effect in comparison to the average magnitude of the overall medium modification to the EEC. First, we observe that this relative directional effect
can reach $\sim 10\%$ at its maximum value for the curve containing the full medium extent (black, $\tau \in (\tau_0, L)$) and for the early stage curve corresponding to the strongest anisotropy profile in Fig.~\ref{fig:qhat} (yellow, $\zeta = 1.5$). Note that the opening angle for which the maximum for EEC$^\Delta$ is attained is different between the black and early stage curves. Further, although not shown explicitly, for the interval $\tau \in (\tau_0, L)$ there is practically no sensitivity to the transition region, i.e., to the value of $\zeta$, while for the colored curves this sensitivity is significant. This is expected since for the early stage time interval, the proportion of time in which the anisotropy is significant is large compared to the full time interval, that contains the EKT data (see bottom panel of Fig~\ref{fig:qhat}). Concerning the sign of EEC$^{\Delta}$, one should note it is not necessarily indicative of the direction of maximal anisotropy. This is a consequence of $\Sigma^\Delta = \Sigma^h-\Sigma^v = \Sigma^h_{\rm med} - \Sigma^v_{\rm med}$ and, thus, a larger suppression along the horizontal clover results in a negative $\Sigma^\Delta$, but a larger enhancement along the vertical clover also results in a negative $\Sigma^\Delta$. However, analyzing the top panel, we see that there is only suppression for the early stage curve, while for the curve with the whole time extent one has a suppression followed by an enhancement. Both of these behaviors are consistent with the bottom panel, such that we can conclude that the magnitude of medium effects is stronger along the horizontal clover, i.e., in the direction of stronger anisotropy as given by the profile in the bottom panel of Fig.~\ref{fig:qhat}. Finally, we note that both sets of curves tend to zero at sufficiently small $\chi$ and, in particular, the black curve does so much faster than the red curve in the top panel. This is a consequence of EEC$^{\Delta} \sim v_2^{\rm EEC} \sim \cO(\chi^3)$, while EEC$_{\rm med} \sim (v_0^{\rm EEC}-v_{0,\,\rm vac}^{\rm EEC}) \sim \cO(\chi)$ (see Eq.~\eqref{eq:v2n_twist_expansion}).

\section{Matter structure effects in a jet quenching Monte-Carlo model}\label{sec:Matter structure effects in a jet quenching Monte-Carlo model}
The previous considerations on the medium induced modifications to jet substructure in an anisotropic medium are based on a simplified analytical model for the directional effects 
and do not account for differences in the jet–medium interactions across the various stages of matter evolution in HICs. Moreover, it is difficult to extend this model calculation beyond leading order in the strong coupling for the computation of energy–energy correlator observables in the HIC context or to include realistic backgrounds. A complementary approach is therefore to study jet dynamics within a dedicated Monte-Carlo framework, which incorporates partial higher-order radiative effects in the jet evolution, albeit typically at the cost of reduced analytic control and a simplified modeling of the underlying jet quenching dynamics.

To this end, we extended the \jewel~\cite{Zapp:2013vla,Zapp:2013zya} medium profile
to include strong directional effects through an anisotropic  temperature profile. The results shown below should therefore be regarded as a qualitative illustration of the effects expected in realistic simulations or in the phenomenological analysis of experimental data.

\begin{figure}[h!]
    \centering
    \includegraphics[width=1\columnwidth]{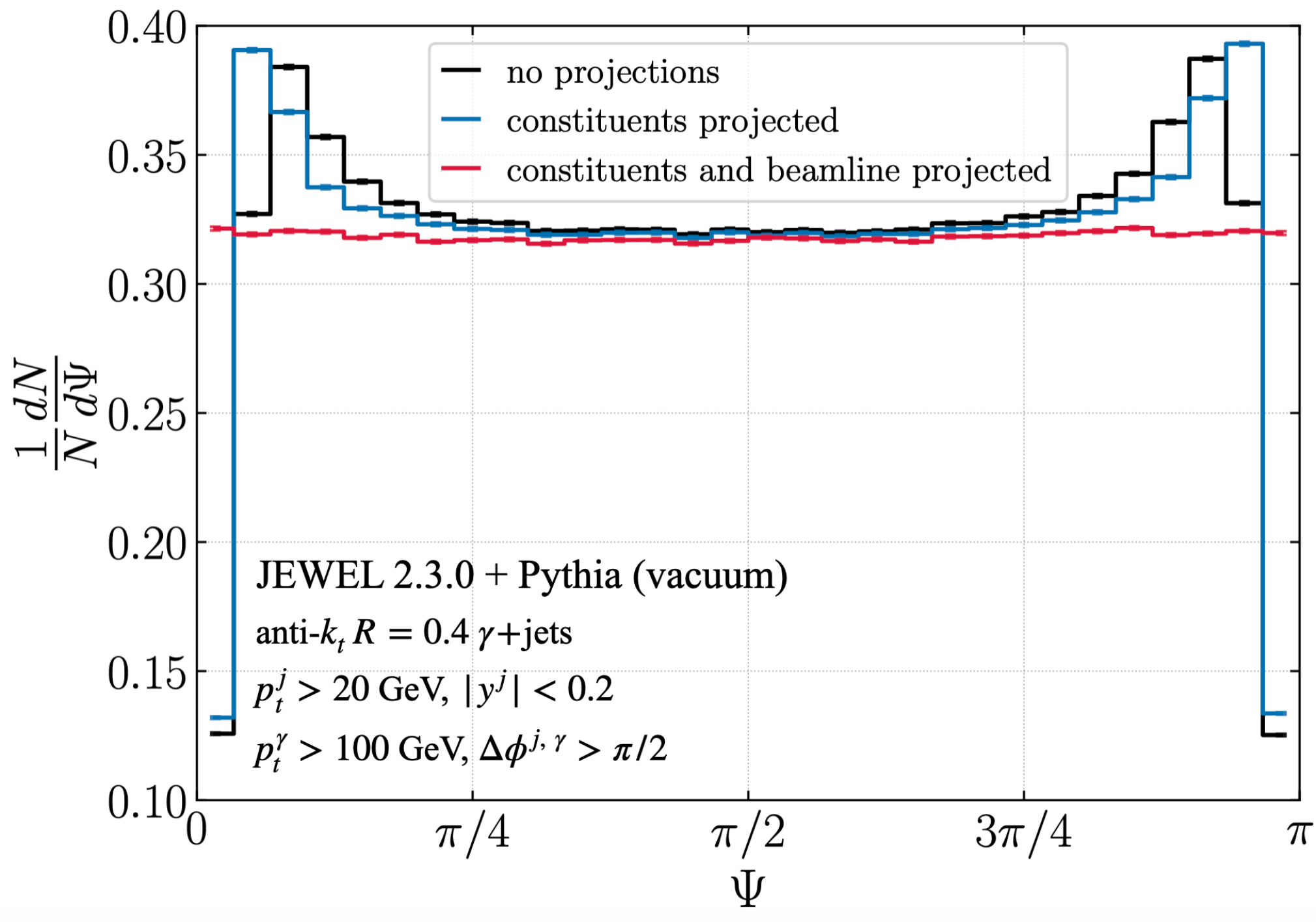}
    \includegraphics[width=1\columnwidth]{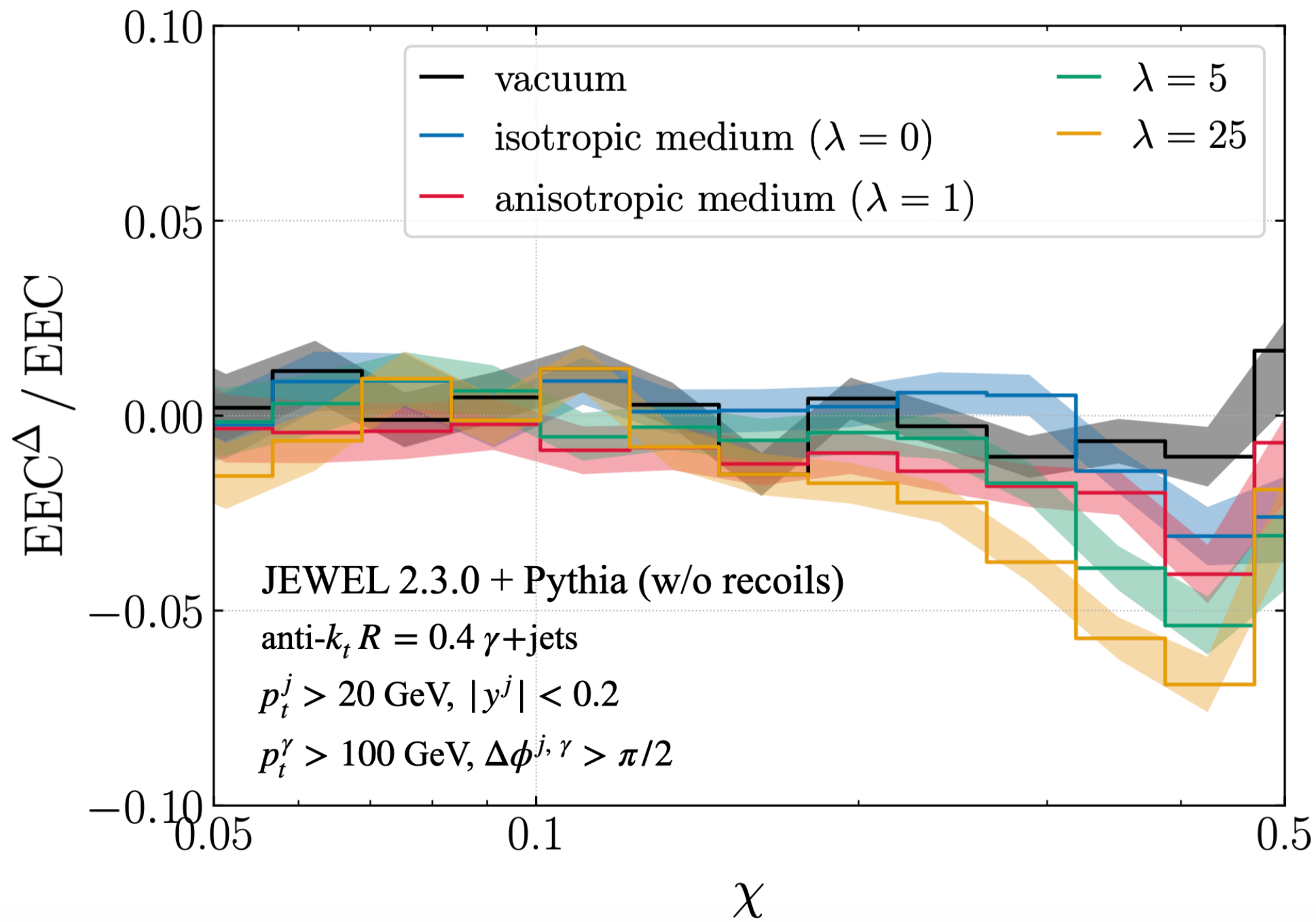}
    \caption{\textbf{Top:} Azimuthal distribution of pairs of jet constituents for vacuum \jewel\ events. The three distributions correspond to different projections of the constituent momenta and of the beamline in the plane transverse to the jet axis, defining different measures of azimuthal angle $\Psi$.  \textbf{Bottom:} Ratio between the \textit{clover} energy-energy correlator (EEC$^{\Delta}$) and the $\Psi$ integrated energy-energy correlator (EEC) inside jets reconstructed from \jewel\ events. The denominator is divided by $4$ to appropriately normalize it to half an hemisphere. The shaded band around each histogram corresponds to the statistical uncertainty.}
    \label{fig:clover_eec_jewel}
\end{figure}

We generated $10^6$ $\gamma+\rm jet$ hadronized events at $\sqrt{s_{NN}} = 5.02$ TeV, with $80$ GeV $ < \hat p_t < 120$ GeV, turning off ISR, MPI and medium recoils. These events correspond to the $[0\%-10\%]$ most central collisions and we set the medium's initial temperature to $T_i = 0.25$ GeV and Bjorken flow initialization time to $\tau_i = 0.6$ fm, keeping the remaining medium parameters to their default values. We reconstructed anti-$k_t$, $R=0.4$ jets at mid-rapidity ($|\eta^j| < 0.2$), with $p_T^j>20$ GeV, produced in the direction opposite to a photon of $p_T^\gamma > 100$ GeV. The remaining kinematic cuts are made explicit in Fig.~\ref{fig:clover_eec_jewel}. The modified medium profile implements a temperature gradient along the beam-axis, such that temperature increases with space-time rapidity $\eta_s$ 
according to a multiplicative factor
\begin{align}\label{eq:F_aniso}
    T(\eta_s) = (1 + \lambda \eta_s^2) T_{\rm default}\,,
\end{align}
where $\lambda = T^{-1}\partial T/\partial \eta_s^2$ at $\eta_s = 0$ controls the strength of the temperature gradients and $T_{\rm default}$ is the default temperature profile in \jewel.\footnote{Note that gradients in the transverse plane $(x,y)$ already exist in \jewel\ due to the initial Woods-Saxon transverse profile. Nevertheless, we have checked that, for mid-rapidity jets, these gradients have a minimal effect in comparison to the $z$-gradients coming from the anisotropy implementation in $\eta_s$.} Naturally, this breaks boost invariance of the underlying background medium. To put the strength of this anisotropy into the context of the medium evolution, we evaluated the temperature gradients along the beam direction near mid-rapidity, $z=0$, at a characteristic hydrodynamization time. This provides a measure of how hydrodynamic the resulting background is, assuming that in the hydrodynamic regime gradients are necessarily small compared to the temperature, while the initial stage dynamics can exhibit strong anisotropies. The leading nonzero gradient reads
\begin{align}\label{eq:d2T}
    \left.\frac{\partial^2 T}{T^3\partial z^2}\right|_{z=0} =  \frac{1}{3}\frac{1+6\lambda}{f^2T_i^2(\tau_it^2)^{2/3}} \, ,
\end{align}
where $f$ characterizes the initial transverse profile of the medium. The characteristic time $t$ can be fixed, for instance, by requiring that $\hat{q}_y/\hat{q}_z$ is sufficiently close to unity. Choosing $t \sim 3$ fm/c for $\lambda \in (0,1,5,25)$, we find that the characteristic second-order gradient values are of order $\sim (0.09,0.6,3,13)$. Thus, for moderately small values of $\lambda$, the modified background relaxes sufficiently close to the hydrodynamic regime, while larger values lead to more pronounced anisotropies that go beyond the validity of a hydrodynamic description.

In Fig.~\ref{fig:clover_eec_jewel}, in the top panel, we show the self-normalized azimuthal distribution of pairs of jet constituents in vacuum events for several measures of the azimuthal angle. 
The red histogram corresponds to the azimuthal angle $\Psi$ defined as the angle between the projections of the momentum difference of two jet constituents and the beamline onto the plane transverse to the jet axis.
For the blue histogram, the azimuthal angle is defined as the angle between the momentum difference of two jet constituents projected onto the plane transverse to the jet axis and the unprojected beam direction taken. Finally, for the black histogram, the azimuthal angle is defined as the angle between the unprojected momentum difference of the jet constituents and the unprojected beam direction.
Naturally, these three definitions of azimuthal angle should coincide at vanishing rapidity and in the strict collinear limit. 
We attribute the peaks in the black and blue curves to effects arising from the underlying spherical geometry at finite rapidity and jet radius.
Note, however, that vacuum effects correlated with the direction of the beamline could also be expected, e.g., initial color connections with beam remnants, as well as geometric distortion due to jets being reconstructed in the $(\eta, \phi)$ space rather than $(\theta, \phi)$. However, we expect these effects to be minimal for mid-rapidity jets, as is supported by the red histogram, where we observe a minimally varying azimuthal distribution.

In the bottom panel of Fig.~\ref{fig:clover_eec_jewel}, we show the EEC$^{\Delta}$ calculated for all pairs of particles inside a jet, with $\Psi$ defined as for the red curve in the top panel. 
This is further divided by the $\Psi$-integrated EEC, which includes both vacuum and medium induced contributions. We first notice that the vacuum (black) band is mostly consistent with EEC$^\Delta \sim 0$, reinforcing that the expected vacuum effects previously alluded to are suppressed for sufficiently small rapidity jets. The increasing trend for $\chi \geq R=0.4$ is due to the finite boundary of the reconstructed jets, and it is common to all bands. Then, for $\lambda=0$ we have default \jewel\ with no added anisotropy (blue band, \textit{isotropic}) and for the remaining bands we show increasingly larger values of $\lambda$, corresponding to larger anisotropies. All bands only significantly deviate from zero for sufficiently large opening angles $\chi$. They exhibit, for most of the large-$\chi$ range, a monotonic behavior with $\lambda$ --- the larger the anisotropy, the stronger the signal in EEC$^\Delta$. Note that for the $\lambda = 0$ case one has EEC$^\Delta \neq 0$, indicating there is some directional effect even in the default \jewel\ configuration. One can verify this by considering Eq.~\eqref{eq:d2T} for $\lambda = 0$, where the second derivative at $z=0$ is non-vanishing. Nevertheless, the points at larger $\lambda$ generate larger EEC$^\Delta$. 
Note that extracting information about the direction of anisotropy from EEC$^{\Delta}$ is, in this case, more involved than in the analytical example shown in Fig.~\ref{fig:clover_eec}.
The reason for this is that, due to the temperature gradient, jet constituents at sufficiently large rapidity are not only more strongly broadened along the beamline, but also subject to enhanced suppression.
This can lead to a larger fraction of the jet’s energy being distributed along the vertical clover, despite the stronger longitudinal broadening.

\section{Discussion and Conclusion}\label{sec:conclusion}
In this work, we have demonstrated that the azimuthal structure of the double-differential EEC in HIC provides a sensitive and theoretically controlled probe of the early, far-from-equilibrium stages of QCD matter. We showed that the celestial OPE of the EEC in the presence of a vector perturbation naturally matches the expected substructure of jets propagating through an anisotropic pre-equilibrium medium, allowing one to potentially isolate directional information associated with the initial state. Focusing on mid-rapidity jets, we argued that hydrodynamic contributions average out, such that a non-vanishing signal in the clover EEC observable directly reflects anisotropies originating from the earliest stages of the collision. We exemplified these observations in an analytical model for jet-medium interactions in the early stages, where the clover EEC naturally couples to the geometry of the background, and carrying a Monte-Carlo study based on a modification of \jewel, where we again observed a direct coupling of the jet to the underlying matter's geometry. Taken together, these results identify azimuthally differential energy correlators as observables with the potential to access the spacetime structure of the pre-hydrodynamic QCD medium.

To conclude this work, we would like to comment on aspects connected to this 
study that may deserve further investigation.

\noindent\textbf{{Higher point energy correlators:}} 
In the main body of this work, we focused on the EEC as a probe of azimuthal structures that can be correlated with early-time properties of the medium.
The reason for this was twofold: on the one hand the EEC is the most reliable observable derived from light-rays operators which might be understood experimentally and theoretically in HIC setups; secondly, the description of branching processes in non-trivial matter backgrounds is poorly understood theoretically, and thus any numerical result would be entirely dominated by modeling.  
Nonetheless, the discussion in the main text can be directly applied to multi-point projected correlators without significant modifications, and it admits a direct extension to fully differential multi-point correlators.
We proceed to illustrate this for the case of the three point correlator (E3C).

\begin{figure}[h!]
    \centering
    \includegraphics[width=.55\columnwidth]{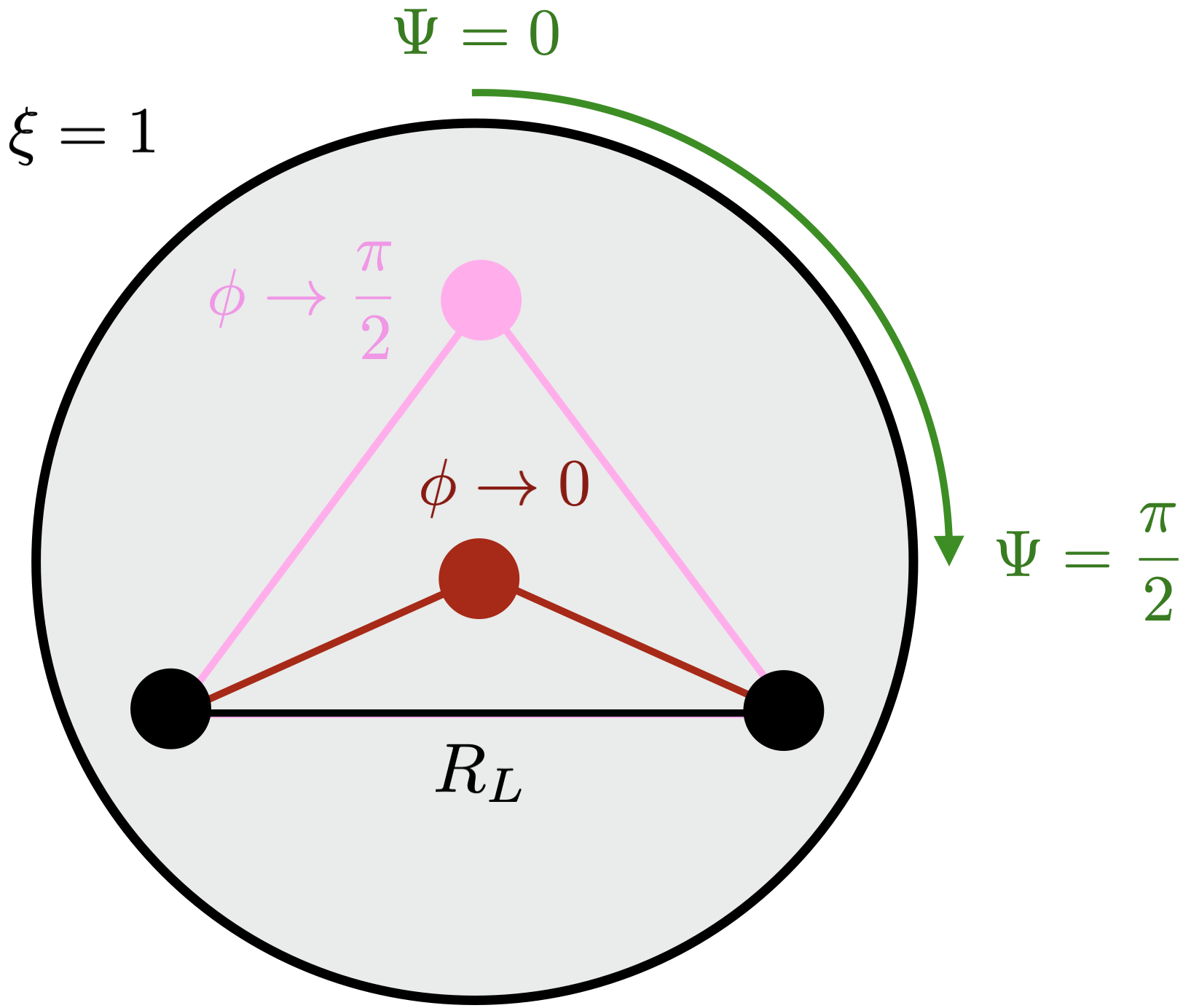}
    \caption{Illustration of the E3C as a way to map the properties of non-trivial matter states. Here we depict that the three detectors can be rotated around ($\Psi=0 \to \pi/2$), while fixing the the length of the triangle side $R_L$. Then varying the position of the central point between equilateral  ($\phi\to \pi/2$) and collapsed ($\phi\to 0$) configurations one can create a \textit{picture} of the underlying matter state; see also~\cite{Caron-Huot:2022lff}. Note that the rotation in $\Psi$ is only relevant when the bulk is not isotropic.}
    \label{fig:E3C}
\end{figure}

In the collinear limit of QCD, the three point correlator is fully determined by the next to leading order splitting function
\begin{align}\label{eq:J3}
 J_3 = \int d\tilde \Phi_3 \, \frac{g^4 \, P_3 \, \,  z_1 z_2 z_3}{2s_{123}^2}\delta\left(1-\sum_i z_i\right)\, ,
\end{align} 
and $s_{ij}= -p_i \cdot p_j$. In general, the above phase space is six dimensional, which is typically reduced to three dimensions, e.g. the length of the triangle edges between the detectors. However, in, e.g. the Glasma, one should further account for both the global placement of the detectors on the sphere and the rotation around the triangle axis, see Fig.~\ref{fig:E3C}. This simplifies in the collinear limit and at mid-rapidity, where the phase space measure reads $d\tilde \Phi_3 = d\Psi/(2\pi )d \Phi_3$, 
where $d\Phi_3$ denotes the measure for the triangle sides and its explicit form can be found in~\cite{Catani:1998nv}.

The decoupling in phase space allows to directly analyze the E3C without requiring an explicit form for the NLO splitting kernel $P_3$. First we recall, that $P_3= P_3(R_L, \phi,\xi, \Psi)$, where $R_L$, $\phi$, and $\xi$ characterize the shape of the correlation function. In particular, in order to extract 
any information about the matter's properties, the large side of the triangle, $R_L$, needs to be taken large enough such that the corresponding EEC is non-trivial, see~\cite{Barata:2025fzd}. More, since the medium imprints itself on the higher twist celestial blocks, it is convenient to take $\xi=1$ and allow $0<\phi<\pi/2$ to vary. This interpolates between a collapsed triangle at $\phi=0$ and an equilateral configuration at $\phi=\pi/2$.\footnote{One should note that this choice of parameters seats at the edge of the applicability of the collinear expansion, where the OPE should matched to the correlator away from the limit, see~\cite{Chen:2022jhb}.} In general, in the limit where only $\phi$ and $\Psi$ are dynamical, we find 
\begin{align}
    P_3 = {\color{blue} P_3^{(0)}} +  P_3^{(2)} \cos(2\Psi) + \cdots  \,, 
\end{align}
by symmetry; the same reasoning is used to derive Eq.~\eqref{eq:EEC_OPE}. Note that each $P^{(i)}$ has a further expansion in twist, which we leave implicit, and the term in blue was first studied in~\cite{Barata:2025fzd}. Again we could construct a \textit{clover} E3C and obtain the shape information in a less differential observable, which we illustrate this in Fig.~\ref{fig:E3C}. Compared to the EEC, here we have now one extra dimension ($\phi$) which allows to further resolve the shape of the early stages. Nonetheless, the form of the celestial block expansion is more involved then for the two-point function and, as far as we are aware, it has not been discussed for the dependence on the triangle rotation, which would be worth exploring.

\noindent\textbf{{Formation time and generalized detectors:}} In the introduction to this work, we argued that the time integrations are effectively cut-off when discussing directional effects since the late time evolution does introduce directional contributions. However, in practice it might still be useful to develop tools sensitive to the dynamics along the longitudinal direction; see~\cite{Apolinario:2020uvt,Apolinario:2024hsm,Apolinario:2017sob} for recent developments along these lines, which can cleanly isolate the dynamics during certain time intervals.

Here, instead of considering phenomenologically driven tools, we discuss constraining the dynamics along the light-ray direction using generalized energy flow operators. These were first considered in~\cite{Korchemsky:2021okt} (see also~\cite{Caron-Huot:2022lff} for related discussion) and can be defined as
\begin{align}
    \mathcal{E}(\omega) = \lim_{r\to \infty}r^2\int_{-\infty}^\infty du \, e^{-i u \omega}   n^i T^{0i}(u+r,nr) \, ,
\end{align}
which matches the standard definition when $\omega =0$. This operator allows to explore the longitudinal structure of the radiation cascade by changing the external frequency $n \omega$, which can be naively associated to that of a probing photon. When acting on a single particle state, this operator transforms the state by injecting energy into it,
\begin{align}
    \mathcal{E}(\omega) |p \rangle \propto\Theta(p^0 + \omega) \delta^{(2)}(\Omega_p - \Omega_n) |p + n\omega \rangle \, ,
\end{align}
which is only possible if $\omega>-p^0$.
However unfeasible such detector operator might be to realize in an experimental setting, it is clear that the sector $\omega>0$ is sensitive to the radiation initiated earlier in time. Thus, if $u \sim t_{\rm form} \sim \tfrac{1}{\chi^2 p_t}$, then taking $\omega\, t_{\rm form}\leq 1$ selects earlier gluon radiation in the jet cascade. As a result, this construction allows to build a longitudinal picture of the fragmentation pattern by tuning $\omega$.

The computation of the EEC with a single generalized operator $\langle \mathcal{E}(\omega) \mathcal{E}(0)\rangle$ at LO in QCD is straightforward, although not particularly illuminating, beyond the previous considerations. Higher order calculations in conformal theories can be found in~\cite{Korchemsky:2021okt}. Here we briefly discuss the behavior of classical terms relevant in the HIC context. Let us consider a particularly instructive case of a spherical shell which is at rest before a time $t_0$ and then flies symmetrically to infinity at constant speed $v$. Notice that introducing angular structure to this setup is straightforward, since it does not affect radial and temporal dependencies. The resulting classical stress energy tensor is given by
\begin{align}
    T^{0i}=\frac{w\gamma v}{r^2}\,\frac{r^i}{r}\,\delta(r-vt)\theta(t-t_0)\,,
\end{align}
where $\gamma$ is just the Lorentz factor and $w$ sets the characteristic energy scale. The time-modified energy flux reads
\begin{align}
    \mathcal{E}(\omega)
    &= w\gamma v\, \lim_{r\to \infty} e^{-i\o r\left(\frac{1}{v}-1\right)}\theta\left(r-v t_0\right)\,.
\end{align}
Thus, the result is sensitive to the order of limits, {\it i.e.} whether one assumes $t_\infty>r_\infty$ or $t_\infty<r_\infty$ and what the limiting values are used. In the EEC discussion, one commonly assumes the former case, allowing all the particles to reach the detector, and in this regime the result is a highly oscillating function (which averages to zero if measured over a finite range in r). In the small $\o$ limit, this expression first becomes non-zero if the radial resolution is sufficiently fine, $\o\ll 1/\Delta R$, and for $\o\ll1/r_\infty$ we smoothly reach the regular case of time-averaged flux.

If instead we consider a single particle, formed at $\r_0$ and $t_0$ with velocity $\v$, then 
\begin{align}
    T^{0i}=m\gamma v^i\,\delta^{(3)}\big(\r-\r_0-\v(t-t_0)\big)\,,
\end{align}
and we readily find
\begin{align}
    \mathcal{E}(\omega)  
    &\simeq m\gamma e^{-i\omega\left(r
\left(\frac{1}{v}-1\right)+\left(t_0-\frac{\r_0\cdot\v}{v^2}\right)+\mathcal{O}\left(\frac{1}{r}\right)\right)}\nn 
&\times\left\{\delta\left(\cos\theta-\frac{v_z}{v}\right)\delta\left(\phi-\arctan\frac{v_x}{v_y}\right)+\mathcal{O}\left(\frac{1}{r}\right)\right\}\,.
\end{align}
Focusing on the Fourier phase, one may set $v=1$ for an energetic particle, in which case the $r$-dependence drops out entirely, while the dependence on the production point remains accessible. Recovering the classical contribution to the correlators $\langle \mathcal{E}(\omega)\mathcal{E}(0)\rangle$ from these two examples is straightforward. It would be interesting to further explore how more intricate matter configurations may be resolved using such operators.

\noindent\textbf{Glasma energy-energy correlations and energy correlators:} It is interesting to notice that the connection between energy correlators and the Glasma state was first raised by Krasnitz and Venugopalan~\cite{Krasnitz:1998ns}. They suggest relating the energy correlation functions extracted from statistical simulations of the Glasma with the energy correlators to measured inside jets. While we, at least partially, address this second point, it is fair to state that a direct connection between our proposed observables and the correlations present in the initial stages of HICs is lacking. To the best of our knowledge, the program suggested by Krasnitz and Venugopalan seems not have been carried out at the time. It would be valuable to explore this direction, especially taking into account the developments in the understanding of the early stages of HICs since then.

\noindent\emph{\textbf{Acknowledgments.}} We are grateful to B. Schenke for a discussion that motivated this work. We would like to thank D. Avramescu and F. Lindenbauer for providing the Glasma and kinetic theory data for the jet quenching parameter, respectively, and for useful comments on the manuscript. We are also grateful to C. Le Roux and K. Zapp for assisting in the implementation of the modified geometry in \jewel, which was based on their work in~\cite{Roux:2024fpv}. We also wish to thank L. Apolinário, D. Avramescu, K. Boguslavsky, J. Brewer, P. Guerrero, M. Kuzmin, K. Lee, T. Lappi, W. Li, F. Lindenbauer, X. M. López and I. Moult for helpful discussions and comments. The work of AVS is supported by Fundação para a Ciência e a Tecnologia (FCT) under contract 2022.06565.CEECIND and by the Basque Government through grant IT1628-22. AVS would also like to acknowledge support from Ikerbasque, Basque Foundation for Science. This work is part of a project that has received funding from the European Research Council(ERC) under the European Union’s Horizon 2020 research and innovation programme (Grant agreement No. 835105, YoctoLHC). JGM acknowledges further support from Fundação para a Ciência e a Tecnologia (FCT), under ERC-PT A-Projects ‘Unveiling’, financed by PRR, NextGenerationEU. JMS has been supported by MCIN/AEI (10.13039/501100011033) and ERDF (grant PID2022-139466NB-C21) and by Consejería de Universidad, Investigación e Innovación, Gobierno de España and Unión Europea – NextGenerationEU under grant AST22 6.5.

\bibliographystyle{JHEP-2modlong.bst}

\bibliography{references.bib}

\clearpage
\onecolumngrid  
\appendix

\section{Details on radiative spectrum in expanding anisotropic matter}
Consider a plasma with a time-dependent, adjoint jet quenching parameter $\hat q(t) = \hat q(t)\Theta(t < L^+)$ and an anisotropy parameterized by an axis dependent quenching ($\hat q_\parallel \neq \hat q_\perp$)~\cite{Hauksson:2023tze}. In the BDMPS-Z formalism, the triple-differential spectrum at large-$N_c$ for the process $g\rightarrow Q\bar Q$ reads~\cite{Barata:2024bqp, Barata:2025uxp}
\begin{align}\label{eq:spectrum_aniso}
    & (2\pi)\frac{d\sigma^{\rm in-in}}{\sigma_0 dzd\chi d\Psi} =    \frac{\alpha_s}{\pi} T_F\,\omega\chi\, 2\text{Re}  \int_{T} -i\frac{\sqrt{c_{1\parallel}}\sqrt{c_{1\perp}}}{\sqrt{c_{3\parallel}}\sqrt{c_{3\perp}}} e^{-i\frac{m_Q^2}{2\omega}\Delta t}\exp\left\{i\frac{\omega\chi^2}{4}\left( \frac{\cos^2\Psi}{c_{3\parallel} }+ \frac{\sin^2\Psi}{c_{3\perp}}\right)\right\}\nn
    & \times \left [\frac{1}{2} P_{qg}(z)\Bigg(c_{1\parallel}+c_{1\perp} +\left(\frac{c_{1\parallel}c_{2\parallel}}{c_{3\parallel}}+\frac{c_{1\perp}c_{2\perp}}{c_{3\perp}}\right) + i\frac{\omega\chi^2}{2}\left(\frac{c_{1\parallel}c_{2\parallel}}{c_{3\parallel}^2}\cos^2\Psi + \frac{c_{1\perp}c_{2\perp}}{c_{3\perp}^2}\sin^2\Psi \right) \Bigg) + \frac{m_Q^2}{4\omega}\right ] \, ,\nn\nn
    & (2\pi)\frac{d\sigma^{\rm in-out}}{\sigma_0 dzd\chi d\Psi} = \frac{\alpha_s}{\pi}T_F\frac{\omega\chi}{\chi^2 + 2(m_Q / \omega)^2}\, 2\text{Re} \int_0^{L^+} dt\, i\frac{\sqrt{c_{1\parallel}}\sqrt{c_{1\perp}}}{\sqrt{c_{2\parallel}}\sqrt{c_{2\perp}}}e^{-i\frac{m_Q^2}{2\omega}\Delta t} \nn 
    & \times \exp\left\{-i\frac{\omega\chi^2}{4}\left(\frac{\cos^2\Psi}{c_{2\parallel}}+\frac{\sin^2\Psi}{c_{2\perp}}\right)\right\} \left[P_{qg}(z)\left(\frac{c_{1\parallel}}{c_{2\parallel}}\cos^2\Psi+\frac{c_{1\perp}}{c_{2\perp}}\sin^2\Psi\right)\chi^2 - i\left(\frac{m_Q}{\omega}\right)^2\right] \, ,\nn\nn
    & (2\pi)\frac{d\sigma^{\rm out-out}}{\sigma_0 dzd\chi d\Psi} = \frac{\alpha_s}{\pi}T_F\frac{\chi}{\left(\chi^2 + 2(m_Q/\omega)^2\right)^2}\left(P_{qg}(z)\chi^2 + 2 \left(\frac{m_Q}{\omega}\right)^2\right) \, ,    
\end{align}
where $\omega = z(1-z)p_0^+$, $P_{qg}(z) = z^2 + (1-z)^2$ is the massless $g \rightarrow q\bar q $ vacuum splitting function stripped of its color factor and $\sigma_0$ is the initial production cross-section. The subscripts $\parallel$ and $\perp$ refer, respectively, to along and orthogonal to some pre-defined direction in the transverse plane. In the main text, this direction is the beamline. The angles are defined as
\begin{align}
    \chi = |\n_1-\n_2|\,,\quad \cos\Psi = \frac{(\n_1-\n_2)\cdot\b}{|\n_1-\n_2|}
\end{align}
where transverse vectors read $\n_i = \p_i/E_i$ and one can write $\n_1 - \n_2 = \P/(z(1-z)E)$, where $\P = (1-z)\p_1 - z\p_{2}$ is the relative transverse momentum of the $q\bar q$ pair. The vector $\b$ is a direction in the transverse plane defining $\Psi = 0$. In the main text, we use $\b$ along the beamline and the transverse plane is defined as the plane transverse to the jet axis, which sits at mid-rapidity. The integration range for the in-in term is $T = \{0 < t < L^+, t < \bar t < L^+\}$. The remaining variables are defined as
\begin{align}
	& c_{1i} = \frac{1}{2iS_{\bar 1 1,i}}\,,\quad c_{2i} = \frac{C_{1\bar 1,i}}{S_{\bar 1 1,i}}\,,\quad  c_{3i}  =  \left(-i\frac{ \hat k_i^2(L,\bar t) (z^2 + (1-z)^2)}{4\omega}\right)-\frac{C_{1\bar 1,i}}{S_{\bar 1 1,i}}\,,\nn
\end{align}
with $\hat k_i^2(L,\bar t) = \int_{\bar t}^L dt\, \hat q_i(t)$, $i=\, 
\parallel,\, \perp$ and the $C_{1\bar 1,i}\equiv C_i(t,\bar t)$ and $S_{\bar 1 1,i}\equiv S_i(\bar t, t)$ functions are determined by
\begin{align}\label{eq:SC_diffeq}
	& \left(\partial_t^2+\Omega_i^2(t)\right)S_i(t,t_0) = 0 \,,\quad S_i(t_0,t_0) = 0\,,\quad \partial_t\left.S_i(t,t_0)\right|_{t=t_0} = 1\,,\nn
	& \left(\partial_t^2+\Omega_i^2(t)\right)C_i(t,t_0) = 0 \,,\quad C_i(t_0,t_0) = 1\,,\quad \partial_t\left.C_i(t,t_0)\right|_{t=t_0} = 0\,,
\end{align}
with the imaginary frequency 
\begin{align}
    \Omega_i = \frac{1-i}{\sqrt{2}}\sqrt{\frac{\hat q_i (z^2 + (1-z)^2}{4\omega}}\,.
\end{align}
These functions further satisfy~\cite{Arnold:2008iy}
\begin{align}
    & C(t_1,t_2) = \partial_{t_2}S(t_2,t_1)\,,\nn
    & S(t_2,t_1) = C(t_1,t_0)S(t_2,t_0) - C(t_2,t_0)S(t_1,t_0)\,,
\end{align}
where the last equality holds for $t_2 > t_1 > t_0$. This means that for $t_2 > t_1 > t_0$ one can write everything in terms of the function determined by the first differential equation in Eq.~\eqref{eq:SC_diffeq}:
\begin{align}
    & S(t_2,t_1) = S(t_1,t_0)\partial_{t_0}S(t_2,t_0)-S(t_2,t_0)\partial_{t_0} S(t_1,t_0) 
\end{align}

\subsection{Harmonic decomposition}
The azimuth-differential EEC can be defined in terms of the triple differential cross-section in Eq.~\eqref{eq:spectrum_aniso} as
\begin{align}
    \frac{d\Sigma}{d\chi d\Psi} = \int_0^1 dz\, z(1-z)\frac{d\sigma}{\sigma_0dzd\chi d\Psi} 
\end{align}
One can do an harmonic decomposition of the differential cross-section as was done in e.g.~\cite{Barata:2024bqp,Barata:2025uxp}, such that the expansion for the EEC reads
\begin{align}\label{eq:harmonic_exp}
   \frac{d\Sigma}{d\chi d\Psi}  = v_0^{\rm EEC}(\chi,\xi)+ \sum_{n=1}^\infty v_{2n}^{\rm EEC}(\chi,\xi) \cos\left(2n \Psi \right)\, , 
\end{align}
where
\begin{align}
    & v_0^{\rm EEC}(\chi,\xi) = \frac{1}{2\pi} \int_0^{2\pi} d\Psi \,\,   \frac{d\Sigma}{d\chi d\Psi} =  \frac{1}{2\pi}\frac{d\Sigma}{d\chi}\, , \nn 
    & v_{2n}^{\rm EEC}(\chi,\xi) = \frac{1}{\pi} \int_0^{2\pi} d\Psi \,\,   \frac{d\Sigma}{d\chi d\Psi} \, \cos\left(2n \Psi \right) \, .
\end{align}  
and $\xi = (\hat q_\parallel - \hat q_\perp)/(\hat q_\parallel + \hat q_\perp)$ characterizes the anisotropy strength. The fact that only even cosine harmonics enter the expansion is due to the invariance of the differential cross-section in Eq.~\eqref{eq:spectrum_aniso} under $\Psi\rightarrow \Psi + \pi$ and $\Psi \rightarrow -\Psi$. Further, this spectrum transforms as $\Psi\rightarrow \pi/2- \Psi$ when $\xi \rightarrow -\xi$, i.e., under exchange of the anisotropy sign one changes the $\parallel$ and $\perp$ axes. This in particular implies that $v_{2n}^{\rm EEC}$ is an odd function of $\xi$ for odd $n$ and an even function for even $n$. 

Let us now inspect the expansion of the harmonics in the anisotropy strength $\xi$ and in the opening angle $\chi$, relevant for the twist expansion discussed in the main text. Starting with the expansion for small $\xi$, first notice that any difference of functions $f(c_{i\parallel})-f(c_{i\perp})$ appearing in the spectrum in Eq.~\eqref{eq:spectrum_aniso} can be expanded in odd powers of $\xi$, while their sum $f(c_{i\parallel})+f(c_{i\perp})$ involves only even powers of $\xi$. The reason for this is that all differences of this form are attached to a $\cos2\Psi$ -- in the $\chi^2$ exponential as well as for the term in the numerator proportional to $\chi^2$ --, and in order to respect the covariance under $\xi \rightarrow -\xi$ in the form $\Psi\rightarrow \pi/2 - \Psi$, only odd powers of $\xi$ can contribute. The same reasoning holds for even powers, which are always attached to $\Psi$-independent terms. Thus, the $\Psi$ dependence of the medium-induced contribution to the total spectrum can be written in the general form
\begin{align}
     \exp\left\{i\frac{\omega\chi^2}{4}\cos2\Psi\sum_{i=0}^\infty a_i\xi^{2i+1}\right\}(b(\xi)+c(\xi)\chi^2+d\chi^2\cos2\Psi\,\sum_{k=0}^\infty e_k\xi^{2k+1})
\end{align}
where $b(\xi)$ and $c(\xi)$ are even functions of $\xi$. If we now further expand the exponential for small $\xi$ we get
\begin{align}\label{eq:expo_expansion}
    \sum_{n=0}^\infty \left(\frac{i\omega\chi^2\xi}{4}\right)^n\frac{1}{n!}\cos^n 2\Psi\left(\sum_{i=0}^\infty a_i\xi^{2i}\right)^n 
\end{align}
Notice that the $n$-th power of $\cos2\Psi$ always comes at least with the $n$-th power of $\xi$. Because $\cos^n 2\Psi$ contains at most the harmonic $\cos2n\Psi$ (no higher), one can conclude that the expansion of coefficients for the $2n$-th harmonic starts at power $\xi^n$. The factor in the numerator does not spoil this reasoning, since it provides either a $\Psi-$independent term or a single additional power of $\cos2\Psi$, multiplied by at least a single factor of $\xi$. Hence, the harmonic that survives at leading order in $\xi$ is the $\cos2\Psi$ and higher harmonics show up necessarily for stronger anisotropies~\cite{Barata:2025uxp}. As for the expansion of the harmonic coefficients in twist (in this context, the expansion in $\chi$), one can draw a similar conclusion. In particular, as a consequence of Eq.~\eqref{eq:expo_expansion}, the $n$-th power of $\cos2\Psi$ is always attached to the $n$-th power of $\chi^2$. By the same reasoning as before, since there is an overall $\chi$ factor in Eq.~\eqref{eq:spectrum_aniso}, one concludes the expansion of the coefficient of the $2n$-th harmonic  starts at power $\chi^{2n+1}$ (twist $\tau = 2n+4$). Note that for the $v_0^{\rm EEC}$ harmonic both the vacuum and medium induced contributions start at power $\chi$. This is a consequence of considering massive final states, otherwise the leading vacuum contribution would be $\chi^{-1}$, i.e, twist $\tau = 2$. This proves Eq.~\eqref{eq:v2n_twist_expansion} -- only the $v_0^{\rm EEC}$ survives at leading twist $\tau = 4$, at subleading twist $\tau = 6$ one has $v_2^{\rm EEC}$ and higher harmonics show up at higher twist.

\section{Comments on the hydrodynamical energy flows}

Working with the classical contribution to Eq.~\eqref{eq:main1}, we write
\begin{align}
&\frac{1}{\sin\chi\sin\Psi}\frac{d\Sigma}{d\chi d\Psi}=\frac{1}{p_t^2}\int\,\sin\theta_d\sin\theta_s\,d\theta_sd\theta_d\,d\phi_sd\phi_d\,\mathcal{E}_1\mathcal{E}_2\notag\\
&\hspace{2cm}\times\delta(\cos\theta_d-\cos\chi)\delta(\cos\phi_d\cos\theta_s\cos\phi_s-\sin\phi_d\sin\phi_s-\cos\Psi)\,,
\end{align}
where $\theta_s$ and $\phi_s$ parametrize $\n_s=\frac{\n_1+\n_2}{|\n_1+\n_2|}$, $\theta_d$ is the relative angle between $\n_1$ and $\n_2$, and $\phi_d$ is an azimuthal angle specifying the orientation around $\n_s$. Resolving the constraints in the collinear limit and assuming that $\mathcal{E}=\mathcal{E}(\theta,\cos\phi)$, {\it i.e.} it is symmetric with respect to mid-rapidity and jet-axis-beam-direction planes (jet axis at $\eta=0$), we further find
\begin{align}
&\frac{1}{\sin\chi}\frac{d\Sigma}{d\chi d\Psi}\simeq\frac{2}{p_t^2}\int\,\sin\theta_s\,d\theta_s\,d\phi_s\,\mathcal{E}\left(\sqrt{\theta_s^2+\frac{\chi^2}{4}+\chi\theta_s\cos(\phi_s-\Psi)},\frac{\theta_s\cos\phi_s+\frac{\chi}{2}\cos\Psi}{\sqrt{\theta_s^2+\frac{\chi^2}{4}+\chi\theta_s\cos(\phi_s-\Psi)}} \right)\notag\\
&\hspace{2cm}\times\mathcal{E}\left(\sqrt{\theta_s^2+\frac{\chi^2}{4}-\chi\theta_s\cos(\phi_s-\Psi)},\frac{\theta_s\cos\phi_s-\frac{\chi}{2}\cos\Psi}{\sqrt{\theta_s^2+\frac{\chi^2}{4}-\chi\theta_s\cos(\phi_s-\Psi)}}\right)\,.
\end{align}
This expression is even under $\Psi\rightarrow\pi-\Psi$ (with $\phi_s\to\pi-\phi_s$) and $\Psi\rightarrow-\Psi$ (with $\phi_s\to-\phi_s$), and, thus, it boils down to a series of positive even powers of $\cos\Psi$, if it exists. Moreover, one may notice that $\Psi$ always enters the azimuthal EEC in a combination $\chi\cos\Psi$, and expanding in powers of $\chi$ it reduces to
\begin{align}
&\frac{1}{\sin\chi}\frac{d\Sigma}{d\chi d\Psi}\simeq \sum_{k=2}\Bigg[\sum^{\tau-4}_{j=0,2,4,\dots}a_{\tau,j}\chi^{\tau-3}\cos(j\Psi)\Bigg]_{\tau=2k}
\end{align}

\end{document}